\def \pc  {\rm pc}
\def \kpc {\rm kpc}
\def \kms {\rm km~s^{-1}}
\def\dmdisp{\langle v^2  \rangle_{\rm   DM}^{1/2} }   

\documentstyle [psfig]{l-aa}
\input psfig

\begin{document}

\thesaurus{4(10.08.1; 10.11.1; 12.04.1)}

\title{ {\it Letter to the Editor}
\\Comment on the dispersion-velocity of galactic dark matter particles}
\author{O. Bienaym\'e$^{1}$, C. Pichon$^2$}
\institute{$^1$Observatoire Astronomique de Strasbourg, 11 rue de l'Universit\'e,
F-67000 Strasbourg, France\\
$^2$Astronomical Institute, University of Basel, Venusstrasse 7,
CH-4102 Binningen, Switzerland}

\offprints{O.Bienaym\'e}

\date{}

\maketitle
\markboth {O.~Bienaym\'e \& C.~Pichon: Comment on the
dispersion-velocity of galactic dark matter}{}

\begin{abstract}

In  a recent Letter,   Cowsik,  Ratnam and Bhattacharjee (1996a)  have
built a  dynamically self-consistent spatial distribution of particles
of galactic dark matter. They have come up  with the rather unorthodox
conclusion that the mean velocity dispersion  of dark matter particles
``should be $600\,\kms$ or larger''.  Their letter triggered immediate
comments (Evans 1996, Gates et al. 1996).  Here we find, as did Cowsik
et al.   (1996a),  that models of  the dark  matter halo  can  be made
consistent with velocity dispersion   much larger than  that  expected
from a  simple  application  of  the   virial  theorem in the    solar
neighbourhood. But in  contrast  to their  conclusions, we  show  that
using their  model, we  also  obtain  solutions with  smaller velocity
dispersion down to $\sim 270\,\kms$.   These more orthodox dispersions
arise because of less constraining boundary conditions for the central
density but do not rely on indirect or model dependent measurements of
the large scale distance behaviour of the rotation curve.
\\

\keywords {Galaxy: halo -- Galaxy: kinematics and dynamics -- dark matter --}
\end{abstract}

We    developed  a numerical    algorithm following  the  prescription
presented by Cowsik et al. (1996a) and we obtain a rotation curve with
the same general    features. This result provides   an  independent
validation of Cowsik et al.'s numerical work. { The distribution of the
baryonic   matter   in   our    Galactic   model  corresponds   to   a
double-exponential disk (with a  scale length of $3.5\,\kpc$,  a scale
height of    $300\,\pc$    and a  surface  density,    $\Sigma_0$,  of
$80~M_\odot\pc^{-3}$ at the solar radius $R_0=~8.5\,\kpc$).  The bulge
's model is  a Hubble profile with a  core radius  $a=~103\,\pc$ and a
central density $\rho_c= 343~M_\odot\pc^{-3}$.}

We check   that  the  velocity   curves  of   these  models  do   tend
asymptotically to $(2/3)^{1/2}\dmdisp =\Theta_{\infty}$.  We determine
velocity curves for  three models with  different  $\dmdisp =350, 450$
and $600\,\kms$  (Fig.~1)  { (the fixed size  of  our rectangular grid
does not  allow us to resolve  simultaneously with sufficient accuracy
both the velocity curve to galactic  radius of $40\,\kpc$ and close to
the galactic    center   below $2\,\kpc$  which   we   have  therefore
ommitted).}

For each dispersion we estimate values of $\rho_{DM}(R_0=8.5\,\kpc,0)$
in order to obtain a flat rotation curve in the range $R=10-30\,\kpc$.
We  note a  dip  at about   $20-25\,\kpc$, which  increases   with the
velocity dispersion (too high  a velocity dispersion prevents a nearly
flat rotation curve).  The velocity curve  maximum near $R=7\,\kpc$ is
produced  by the exponential disk of  visible  matter.  At much larger
galactic radius  $R>40\,\kpc$  the dark matter contribution  dominates
the velocity curve  which tends towards $\Theta_\infty$.  For instance
in Fig.   1, the  lowest curve  (3) that   corresponds to the  largest
$\dmdisp$ reaches $490\,\kms$  at large $R$  while curves (2)  and (1)
only  reach   367  and   $286\,\kms$.    Increasing $\rho_{DM}(R_0,0)$
increases the mass contribution of the dark halo in the central region
and convergence towards   $\Theta_{\infty}$ is reached more   rapidly.
Yet, in this  instance, the velocity  curve does not remain flat below
$40\,\kpc$.\\ The  most  unusual  feature of  the  models  proposed by
Cowsik et al.  is that the central dark matter  density is so low that
the visible matter dominates  the   velocity curve at  small  galactic
radius and allows for a galactic population with an extension stopping
at    $30-40\,\kpc$     while     its      velocity   dispersion    is
$(2/3)^{1/2}220=180\,\kms$  (very  similar to   that  of  the Globular
Cluster population for instance).  The dark matter population having a
larger  $\dmdisp$  dominates fully after  $30\,\kpc$  and  as shown in
Cowsik's  reply (1996bc)  this   dark matter  component   may  stop at
$100\,\kpc$ without significant effect at small galactic radius.

\begin{figure*}
\psfig{figure=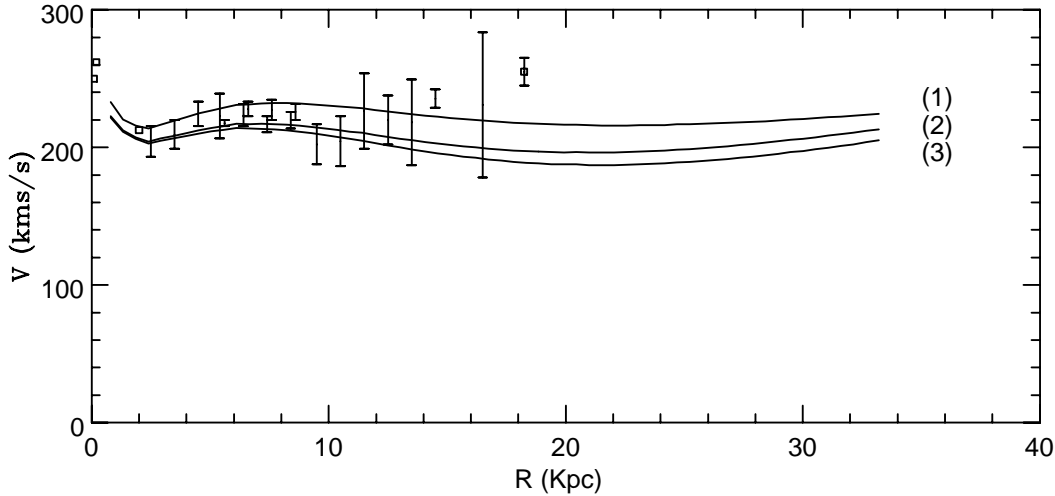,width=15.cm,bbllx=15pt,bblly=160pt,bburx=590pt,bbury=500pt,clip=true}
\caption{ Flat rotation curve of the Galaxy  $\dmdisp=350$ (1),
$450$ (2) and $600\,\kms$ (3).  { The observational data and 
their error bars
are  those given by   Cowisk et al   :  from Fich  et al  (1989), below
$2\,\kpc$ from Burton \& Gordon (1978)  and for $R>17\,\kpc$ from Fich
\& Tremaine (1991). The corresponding $\chi^2$  take   respectively
the values of  $14$, $18$  and $22\,\kms$  for
models (1), (2) and (3)}}
 \end{figure*}

Cowsik's   results are in  conflict with  available  data only  in the range
$20-40\,\kpc$. In this range, constraints arise from the dynamics of distant
objects  such as globular clusters. These  give  access to the galactic mass
inside   $R\sim 40\,\kpc$, and suggest  that  its distribution is consistent
with a flat rotation curve up to that distance (Dauphole  \& Colin 1996). In
this   reference, input   data are  6D  ($\vec   r, \vec v$)  positions  and
velocities for a critical subset of  globular clusters, giving for the first
time strong  constraints on the  rotation curve   in  the analysed range  of
distances. Their most probable fit corresponds to a flat rotation curve, but
no  upper limits  to  a rising rotation   curve are  given. Of  course, such
results remain model dependent  and do not  exclude a rising (or decreasing)
rotation curve beyond $40\,\kpc$.

We remark  that  the data  used  by Cowsik et  al.  (1996a) correspond  to a
rotation curve determined with the assumption that the solar galactic radius
is $R_0=8.5\,\kpc$ (a value  recommended by the  IAU - see Kerr \& Lynden-Bell
1986), used to  ``facilitate intercomparison  of the  work  of the different
authors''   (Fich \& Tremaine  1991).  Smaller  values  (Reid 1993)  are now
generally accepted and  lead to a flat  or locally decreasing rotation curve
(Fich \& Tremaine 1991).

Our  models do  not take   as data  input the  mass  determination from   17
satellites  of  the   Galaxy   more   distant than  $50\,\kpc$   (Kochaneck,
1996). Only 3 have   complete ($\vec  r,  \vec  v$) data and  mass  estimate
depends on complementary hypotheses    like  velocity isotropy or     radial
gradient    of the density.   Similarly we   do  not incorporate the timing
argument as means to estimate the  total mass of the  Galaxy since as argued
by  Kochaneck (1996) `` The  classical Local Group  timing model of Kahn and
Woltjer 1959 assumes that the orbits are radial and provides lower bounds on
the mass.''

Within the framework of our isothermal model, we find that an upper limit of
$600\,\kms$   is    consistent with  a flat  rotation    curve in  the range
$R=10-40\,\kpc$. A  range    of  possible  galactic  dark   matter  velocity
dispersions from 270   to $600\,\kms$ are   shown to be also  consistent. We
remark that the  velocity dispersion  can  be much larger beyond  $40\,\kpc$
without   significant measurable effects on  the   rotation curve below this
radius. If the velocity curve of the Galaxy is only well defined over such a
small  distance  interval, it   is  difficult to   model  it  with a  unique
isothermal  model.  One should  therefore be careful  in estimating velocity
dispersions from a crude application of the  virial theorem, as mentioned by
Cowsik et al. (1996a).\\

We find a large range of possible $\dmdisp$ with  dispersion as low as
$270\,\kms$ if we consider the  rotation curve used  by Cowsik et al.,
and from $270$ to $600\,\kms$ if we consider  a flat rotation curve up
to $40\,\kpc$. The   discrepancies between Cowsik's et  al. conclusion
and ours arises from the   fact that we did  not  fix the dark  matter
density   at   the origin    $\rho_{D.M.}(0,0)$.   Cowsik et  al.  set
$\rho_{D.M.}(0,0)=1\,Gev/cm^3$ constant,  arguing   that the dynamical
measure  of   the mass density in  the    solar neighbourhood is $\sim
0.3\,Gev/cm^3$.  This value of $\rho_{D.M.}(R_0,0)\sim  0.3\,Gev/cm^3=
0.008\,M_\odot   /pc^3$   is    very  different   from    the   values
($0.05-0.1\,M_\odot /pc^3$) obtained in references (Oort 1960, Bahcall
1984) cited by Cowsik et al. (1996a). The  exact estimate of the local
dynamical   mass     density  $\rho_{Dyn.}(R_0,0)$   remains   a  very
controversial subject, though it is now determined with smaller errors
(see review by   Cr\'ez\'e 1991 \& Kuijken  1995).   Models plotted in
Fig.~1   have small   $\rho_{D.M.}(R_0,0)$  compatible  with this last
determinations   of $\rho_{Dyn.}(R_0,0)$;  these   models are in rough
agreement with the data  used by Cowsik  in the range $R<20\,\kpc$ and
have much smaller $\dmdisp$  values than those  obtained by  Cowsik et
al. (1996a).\\

Cowsik et al. have pointed out the important fact that models of dark matter
halo can  be made consistent with velocity  dispersion much larger than that
expected from  a  simple application of   the virial  theorem in  the  solar
neighbourhood.  Their model should  be   applied to  existing 6D  data  from
globular clusters in  order to obtain a  realistic upper limit on $\dmdisp$.
Here  we have shown that,  in contrast to   Cowsik et al.  assertion,
models with  small orthodox $\dmdisp  \sim 270\,\kms$ are also  in agreement
with their own observational constraints.   Our conclusions suggest that the
efforts on the  part of experimentalists should  not be directed towards the
search for very hypothetical  high velocity dispersion dark matter particles.
These effort  should on the contrary be  directed towards better astrometric
observations in order to reach valuable absolute proper motions for the most
distant  galactic satellites and globular clusters  that may be achieved for
instance with high resolution observations with HST or adaptative optics.

\vskip 0.25cm

\noindent {\it CP aknowledges funding from the Swiss NF.}

\end{document}